# Magnetocaloric effect and Magnetothermopower in the room temperature ferromagnet $Pr_{0.6}Sr_{0.4}MnO_3$

D. V. Maheshwar Repaka, T. S. Tripathi, M. Aparnadevi and R. Mahendiran[1]*

Department of Physics, 2 Science Drive 3, Faculty of Science,

National University of Singapore, Singapore 117542, Singapore

**Abstract**

We have investigated magnetization($M$), magnetocaloric effect(MCE) and magnetothermopower(MTEP) in polycrystalline $Pr_{0.6}Sr_{0.4}MnO_3$, which shows a second-order paramagnetic to ferromagnetic transition near room temperature ($T_C$ = 305 K). However, field-cooled $M(T)$ within the long range ferromagnetic state shows an abrupt decrease at $T_S$ = 86 K for $H$ < 3 T. The low temperature transition is first-order in nature as suggested by the hysteresis in $M(T)$ and exothermic/endothermic peaks in differential thermal analysis for cooling and warming cycles. The anomaly at $T_S$ is attributed to a structural transition from orthorhombic to monoclinic phase. The magnetic entropy change ($\Delta S_m = S_m(H)-S_m(0)$) shows a negative peak at $T_C$ (normal MCE) and a positive spike (inverse MCE) at $T_S$. $\Delta S_m$ = -2.185 J/kg.K (-3.416 J/kg.K) with refrigeration capacity RC = 43.4 J/kg (103.324 J/kg) for field change of $\Delta H$ =1.5 T (3 T) at $T_C$ = 304 K is one of the largest values reported in manganites near room temperature. Thermopower ($Q$) is negative from 350 K to 20 K, shows a rapid decrease at $T_C$ and a small cusp around $T_S$ in zero field. The MTEP [$\Delta Q/Q(0)$] reaches a maximum value of 25% for $\Delta H$ = 3 T around $T_C$ which is much higher than 15% dc magnetoresistance for the same field change. A linear relation between MTEP and magnetoresistance and between $\Delta S_m$ and $\Delta Q$ are found near $T_C$. Further, ac magnetotransport in low dc magnetic fields (H ≤ 1 kOe), critical analysis of the paramagnetic to ferromagnetic transition and scaling behavior of $\Delta S_m$ versus a reduced temperature under different magnetic fields are also reported.

---

[1] Author for correspondence (E-mail:phyrm@nus.edu.sg)

Coexistence of large magnetic entropy change and magnetothermopower around room temperature makes this compound interesting for applications.

## I. Introduction

Perovskite manganites having the general formula $R_{1-x}A_xMnO_3$ (R is a trivalent rare earth cation and A is divalent alkaline earth cation) have been extensively investigated during the last two decades due to colossal magnetoresistance and electroresistance effects exhibited by them, and rich physics behind their novel phase diagrams. In recent years, much attention has been paid to another exciting property of manganites known as the magnetocaloric effect (MCE).[1,2] The MCE refers to changes in adiabatic temperature ($\Delta T_{ad}$) or isothermal magnetic entropy ($\Delta S_m$) of a magnetic material upon magnetization and demagnetization. The MCE is attractive because it is the working principle of the emerging technology called magnetic refrigeration, which is considered to be energy efficient and environmentally friendly compared to the conventional vapor compression based refrigeration. The magnetic entropy change ($\Delta S_m$) is dependent on the rate of change of magnetization ($M$) with temperature and also on the heat capacity ($C$) through the Maxwell's relation $\Delta S_m = \int_0^H \frac{1}{C(T,H)} \left(\frac{\partial M}{\partial T}\right)_H dH$. Due to the rapid change of $M(T)$ at $T_C$ during first–order transition, MCE is greatly enhanced in materials exhibiting first-order magneto-structural transition such as $Gd_5(Si_2Ge_2)$ [3], $Le(FeSi_{1-x})_{13}$,[4] $MnFe(P,Si,Ge)$ etc.[5] However, hysteresis associated with the structural phase transition, mechanical instability and cost of the raw materials pose challenges that have to be overcome for practical applications of these compounds. Meanwhile, researchers are looking to synthesize and characterize new materials with second-order phase transition showing $\Delta S_m$ value comparable to or larger than the metallic gadolinium, which shows excellent magnetocaloric effect ($\Delta T$ = 17 K change in a field of $\Delta H$ = 7 Tesla[6] and 5.6 K for $\Delta H$ = 2 T[7]) near room temperature ($T_C$ = 293 K) but its high cost, oxidization and brittle nature are matters of concerns for large scale device applications.

A first-order magnetic phase transition is often accompanied by abrupt change in volume without a change in crystallographic symmetry or structural phase transition. The change in lattice entropy ($\Delta S_{latt}$) can have the same or opposite sign as the magnetic entropy change ($\Delta S_m$) and hence it can either boost or reduce the total measured entropy change ($\Delta S = \Delta S_m + \Delta S_{latt}$). For example, $\Delta S_{latt}$ in $Ni_{45.2}Mn_{36.7}In_{13}Co_{5.1}$ alloy is opposite in sign to the $\Delta S_m$ and in $Gd_5S_2Ge_2$, lattice entropy accounts for more than 50% of the total MCE.[8] The electronic part of entropy change is generally assumed to be small.

Interestingly, both first-order and second-order phase transitions occur in the $Pr_{1-x}Sr_xMnO_3$ series as a function of temperature or composition (x) [9,10] and hence it is an interesting series to investigate the influence of structural transition on $\Delta S_m$. While the compounds $0.2 \leq x \leq 0.45$ show a second order paramagnetic (PM) to ferromagnetic (FM) transition, the half doped compound x = 0.5 shows a second-order PM to FM transition at $T_C$ = 260 K followed by a first order FM to antiferromagnetic (AFM) transition at a lower temperature ($T_N$ = 125 K). This FM to AFM transition is also accompanied by tetragonal to monoclinic phase transition while there is no structural symmetry change across the PM to FM transition.[11] Bingham et al.[12] reported normal magnetocaloric effect around $T_C$ and inverse MCE around $T_N$ in x = 0.5 On the other hand, x = 0.54 compound shows a first-order phase transition from PM to AFM state around $T_N$ = 210 K without the intermediate FM state. The PM to AFM transition is accompanied by tetragonal to orthorhombic structural transition.[9] A large inverse MCE and field-induced structural transition were demonstrated in this compound.[13] In contrast to the above two compounds, x = 0.4 is a room temperature FM and it shows an orthorhombic to a monoclinic structural transition around $T_S$ = 90 K much below the FM transition. [14] It is of our interest to investigate MCE around the Curie temperature which happens to be around room temperature ($T_C$ = 305 K) in $Pr_{0.6}Sr_{0.4}MnO_3$ and also how the MCE is affected by the low temperature structural transition. We also investigate the temperature and magnetic field dependences of thermo electric power (TEP) in this compound. Although temperature dependence of the TEP in zero field was reported for $0.48 \leq x \leq 0.6$ in the $Pr_{1-x}Sr_xMnO_3$ series,[15] neither TEP in zero field nor effect of magnetic field on TEP has been reported in x = 0.4 so far. In addition, we also report frequency dependent electrical transport in small dc magnetic fields ($H$ = 0 to 1 kOe). Temperature and field dependences of four probe ac electrical transport in metallic or low resistivity have been seldom reported compared to dielectric studies in insulating manganites.[16], Recently, we have reported the occurrence of more than 40% ac magnetoresistance in a low magnetic field of $H$ = 1 kOe in $La_{0.7}Sr_{0.3}MnO_3$ [17]and $La_{0.7}Ba_{0.3}MnO_3$.[18] From this perspective, it is interesting to check whether $Pr_{0.6}Sr_{0.4}MnO_3$ also show colossal ac magnetoresistance. Finally, we also analyze critical exponent associated with the PM-FM transition and show scaling of $\Delta S_m$ with magnetic field and temperature.

## II. Experimental details:

Polycrystalline sample of $Pr_{0.6}Sr_{0.4}MnO_3$ was prepared by the solid state reaction method. Powders of $Pr_6O_{11}$, $SrCoO_3$ and $Mn_3O_4$ were mixed in proper molar fraction. After initial mixing and grinding the compound was annealed at 1000ºC, 1100ºC and 1200ºC each for 12 hours and subsequently final sintering of a pellet was carried out at 1200 ºC for 24 hours. Room temperature powder x-ray diffraction confirmed that the sample is single phase with orthorhombic crystal structure (Pnma space group). Direct current (DC) Magnetoresistance and Magnetization were measured using a Physical Property Measuring System (Quantum Design Inc., USA) equipped with vibrating sample magnetometer probe. The thermoelectric power (TEP) was measured between 365 K down to 10 K and using an automated homemade setup that is interfaced to the PPMS. The PPMS provides a platform to vary temperature and magnetic field. For TEP measurement, a rectangular sample was mounted between two copper blocks and two chip resistors were used as heat source. In this method, at a stabilized base temperature of the cryostat, a small temperature gradient ($\Delta T = 1$ K) is generated across the sample length, and the thermoelectric voltage is recorded using copper leads. The temperature difference was measured using a Chromel-Constantan differential thermocouple after steady state was reached. The spurious and offset voltages of the measuring circuit were eliminated by reversing the temperature gradient and averaging the recorded voltages. The apparatus was tested for accuracy by measuring $Q$ on a thin piece of pure lead with respect to copper[19]. The applied magnetic field was perpendicular to the direction of the dc current and length of the sample in both magnetoresistance and magnetothermopower measurements. Four probe ac electrical impedance ($Z = R+iX$) as a function of frequency ($f = $ 100 kHz- 10 MHz), temperature and magnetic field was measured using an Agilent 4294A impedance analyzer and PPMS.

## III. Results and Discussion

### (a) Magnetization and the critical behavior

Fig. 1 shows the temperature dependence of dc magnetization (*M*) while cooling and warming under different dc magnetic fields from $H = 0$ to 3 T. The sample undergoes a PM to FM transition around $T_C = 305$ K as indicated by the rapid increase of $M(T)$ under H = 1 kOe. While cooling, smooth increase of $M(T)$ below $T_C$ is interrupted by a step-like decrease at $T_S = 86$ K within the ferromagnetic state. While warming, $M(T)$ shows an abrupt increase at 98 K with a pronounced hysteresis at around $T_S$. The feature at $T_S$ does not shift down in temperature with increasing $H$ but the width of the hysteresis vanishes for $H > 1$ T. It suggests that the step-like decrease observed while cooling is not due to AFM transition. To characterize the low temperature anomaly, we carried out differential thermal analysis (DTA) which is shown in the inset (a) of fig.1. The DTA technique makes use of two Pt-100 resistance thermometers connected in differential mode. The sample is placed on one of the thermometers while the other one is used as a reference. The temperature differences between these two Pt thermometers is read as a function of magnetic field when the base temperature changes.[20] The DTA in our sample shows an exothermic peak at $T$ = 89 K and endothermic peak at T = 100K while cooling and warming, which confirms the 1st order nature of low temperature transition. We attribute the anomaly at $T_S$ is due to orthorhombic to monoclinic structural phase transition upon cooling as suggested by neutron diffraction study on a similar composition by C. Ritter *et al.*[21] Ritter *et al.* also found that the structural transition is not complete even at 1.6 K. Their refinement of neutron diffraction data gave 88% monoclinic (*I2/a*) and 12% orthorhombic (*Pnma*) phases at 1.6 K. C. Boujleben *et al.*[22] determined coexistence of 73% monoclinic and 27% orthorhombic phases in their sample from neutron diffraction study.

The main panel of Fig. 2 shows magnetization isotherms at $T$ = 10 and 120 K. The $M(H)$ behavior at 10 K is typical behavior of a ferromagnet. Interestingly, the values of $M(H)$ for $H < 1$T at 10 K are lower than for $T$ = 120 K, which is consistent with the decrease of $M(T)$ found below $T_S$. The inset (a) on the left shows $M(H)$ isotherms measured from 10 K to 350 K in a temperature interval of 5 K. These data were later used to calculate the magnetic entropy change.

To understand the order of magnetic phase transition and critical behavior by using scaling hypothesis we have taken $M(H)$ isotherms at each 2 K difference from 294 K to 320 K (see Fig. 3(a)). The slope of $M^2$ Vs H/M curves (Arrot plots) can determine the order of phase transition. The positive slope indicates a second -order transition while a negative slope corresponds to first- order transition. [23] The positive slope of $M^2$ vs H/M curves above and below $T_C$ confirms that the high temperature PM-FM transition is of second order in nature. According to the scaling hypothesis, the critical behavior a magnetic systems showing a second-order magnetic phase transition near the Curie temperature can be characterized by a set of critical exponents which are interrelated.[24] The critical exponents associated with the spontaneous magnetization ($M_s$), inverse susceptibility ($\chi^{-1}$) and magnetization isotherms at $T_C$ were calculated by fitting the experimental isotherms using the following scaling relations for second-order phase transitions:[25][26]

$$M_S(T) = M_0(-\varepsilon)^\beta, \quad \varepsilon < 0, T < T_C \quad (1)$$

$$\chi_0^{-1}(T) = (\frac{h_0}{M_0})\varepsilon^\gamma \quad \varepsilon > 0, T > T_C \quad (2)$$

$$M = DH^{1/\delta} \quad \varepsilon = 0, T = T_C \quad (3)$$

Where $\varepsilon = (T - T_C)/T_C$ is the reduced temperature and $M_0$, $(h_0/M_0)$ and $D$ are the critical amplitudes. β, γ and δ are critical exponents associated with $M_S$, $\chi_0$ and $T_C$, respectively. Generally, the critical exponents can be determined by analyzing the Arrot plot at temperature around the critical point. According to Arrot-Noakes equation of state $(H/M)^{1/\gamma} = (T - T_C)/T_C + (M/M_1)^{1/\beta}$, where $M_1$ is a material constant. Hence, in the mean field mode, the Arrot plot drawn as $M^2$ versus $H/M$ curves should be a series of parallel straight lines in the high field range, and the line at $T_C$ should pass through the origin with β = 0.5, γ = 1. However, the curves in the Arrott plot shown in fig.3 (b) are non-linear which indicates that the critical exponents (β = 0.5, γ = 1) based on the Landau mean-field theory alone cannot explain the critical behavior of this compound. The modified Arrott plot with Heisenberg critical exponents with β = 0.365, γ = 1.336 was tired out and it results in parallel straight lines as shown in fig.3 (c). By extrapolating the modified Arrott plots from the high field region to $(H/M)^{1/\gamma} = 0$ for $T < T_C$ and $(M)^{1/\beta} = 0$ for $T > T_C$, the spontaneous magnetization $M_s(T,0)$ and the inverse initial susceptibility $\chi_0^{-1}(T,0)$ were calculated for every straight line and they are plotted in Fig. 3(d). We found β = 0.364 and γ = 1.336 with $T_C$ = 304.6 K. According to the statistical theory, these three exponents must fulfill the Widom Scaling relation $\delta = 1 + \frac{\gamma}{\beta}$, which should be equal to 4.66 at $T = T_c$. Experimentally, the exponent δ is determined by fitting the isotherm magnetization curves at $T_C$ and it is found to be δ = 4.65 which is very close to the expected value.

**(b) Magnetocaloric effect**

Isothermal M-H curves were used to estimate the magnetic entropy change ($\Delta S_m$) with the help of the numerical approximation to the Maxwell relation. Fig. 4 shows the change of magnetic entropy $(-\Delta S_m)$ as a function of temperature under different dc magnetic fields. The $-\Delta S_m$ increases with lowering temperature and shows a positive peak at $T_C$ below which it decreases down to 100 K. Then, it shows a dip ($-\Delta S_m$ is negative) around $T_S$ and then gradually increases towards zero with further lowering temperature. Hence, the compound exhibits both normal ($-\Delta S_m$ is positive) and inverse ($-\Delta S_m$ is negative) magnetocaloric effects. The $\Delta S_m$ vs $T$ curves are symmetric about $T_C$ as expected for a 2$^{nd}$ order FM transition and the position of the $-\Delta S_m$ peak at $T_C$ is not affected with increasing magnetic field. The amplitude of the $-\Delta S_m$ peak increases from 1.648 J/kg,K for a field change of $\Delta H = 1$ T to 4.65 J/kg.K for $\Delta H = 5$ T. However, the amplitude of the inverse MCE at $T_S$ is weakly dependent on $H$ and is about -0.578 J/Kg K. Besides a large $-\Delta S_m$, refrigeration capacity (RC) has to be large for the material to be suitable for practical application. The RC defined as $RC = \int_{T_1}^{T_2} \Delta S_m(T) dT$ is an important technical quantity which quantifies heat transfer between the hot reservoir at temperature $T_1$ and cold reservoir at temperature $T_2$ in an ideal Carnot cycle. $T_1$ and $T_2$ are taken as the extreme temperatures corresponding to full width at half-maximum of $\Delta S_m$ for each field. The values of $-\Delta S_m$ and RC are plotted as a function of magnetic field in the inset (a) of fig.4. The RC increases from 27.5 to 182.5 J/kg as the changes from $\Delta H = 1$ to 5 T. For $\Delta H =1.5$ T, estimated values $-\Delta S_m = 2.185$ and RC = 43.4 J/Kg, which are slightly higher than the values reported for $La_{0.7-x}Pr_xSr_{0.3}MnO_3$[27]

Oesterreicher and Parker[28] using the mean-field approach derived a proportional relation between the field dependent change in the magnetic entropy ($\Delta S_m$) and applied field ($H$). At $T = T_C$, the magnetic entropy change was found to vary as $\Delta S_m \approx -1.07 q R \left( \dfrac{g \mu_B J H}{k_B T} \right)^{2/3}$, where $q$ is the number of magnetic ions per mole, $R$ is the gas constant, and $J$ is the total angular momentum. Above $T_C$, $\Delta S_m$ is quadratic in magnetic field ($\Delta S_m = -\frac{qR}{6}\left(\frac{p_e \mu_B H}{k_B(T-\theta)}\right)^2$ where, $p_e$ is the effective magnetic moment in the paramagnetic state and $\theta$ is the paramagnetic Curie temperature obtained from the inverse susceitpibility . Recently, Franco et. al.[29] showed that $\Delta S_m$ in a second- order PM-FM phase transition can be expressed as $\Delta S_m \alpha H^n$, for $T < T_C$ if a proper temperature scaling is introduced. They showed that the plots of the magnetic entropy change normalized to it is maximum value ($\Delta S_m/\Delta S_{Max}$) for different magnetic fields versus a reduced temperature $\theta = \frac{T-T_p}{T_r - T_p}$ falls on to a single master curve. Here, $T_p$ and $T_r$ are temperatures corresponding to the maximum magnetic entropy change ( $\Delta S_{Max}$ ) and half the maximum value ($\Delta S_{Max}/2$), respectively. In insets of Fig. 5, we plot -$\Delta S_m$ vs $H^n$ plot for the three regions of $T >> T_C$, $T \sim T_C$ and $T << T_C$ corresponding to $n$ = 2, 2/3 and 1 respectively. The anticipated field dependence is mostly obeyed. The main panel of Fig. 5 shows the value of $n$ extracted from the $\Delta S_m$ versus $H$ curves at different temperatures. It is seen that while $n$ = 2 far above $T_C$ and $n$ = 2/3 below $T_C$ , it goes through a minimum value of $n$ = 1 close to the $T_C$, as predicted by Franco *et al*. Fig. 6 shows $\Delta S_m/\Delta S_{Max}$ versus the reduced temperature $\theta$ for different magnetic fields. It is found that the data for all the magnetic fields collapse into a single master curve.

(C). **Direct (dc) and alternating current (ac) electrical Transport**

Now let us discuss the dc and ac electrical transport. Fig. 7 shows the dc electrical resistivity as a function of temperature under $H$ = 0 and 3 T magnetic fields. While lowering temperature, $\rho(T)$ shows insulating ($\partial \rho / \partial T < 0$) behavior in the paramagnetic state. It shows a small kink around $T$ = 305 K ($\approx T_C$) and but goes through maximum around $T_p$ = 213K. The departure of $T_p$ from $T_C$ is most likely due to the presence of high resistive grain boundaries.[30] As the temperature decreases, magnetization of ferromagnetic grains increase and the resistivity falls below $T_p$ once the percolation threshold for metallic conduction is reached. The structural transition at temperature $T_S$ hardly affects the resistivity except for introducing a slight slope change. When a magnetic field of $H$ = 3 T is applied, the kink at $T_C$ is suppressed and, $T_p$ shifts by 9 K and magnitude of the resistivity below $T_C$ decreases. The dc magnetoresistance, MR = [ρ(0)- $\rho(3T)]/\rho(0T)$ is shown in right scale of Fig. 7. The dc MR reaches ≈-15% at the $T_C$ and increases up to -39% at 10 K. The structural transition at temperature $T_S$ shows up as a slight slope change in the resistivity. It is known that electrical conduction in the paramagnetic insulating state of manganites, in general, is dominated by adiabatic hopping of small polarons, which obeys the relation $\rho(T) = \rho_0 T \exp(E_p/k_B T)$ where $E_p$ is the activation energy for the electrical transport and $k_B$ is the Boltzmann constant. $E_p = W_H + E_s$, where $E_s$ is the activation energy for thermopower and $W_H$ is the binding energy for polaron.[31] The plot of $\ln(\rho/T)$ versus $1/T$ (see the inset) above $T_C$ shows a linear dependence which suggests that thermally activated polaronic conduction dominates the charge transport in the paramagnetic state in our sample. The estimated activation energy is $E_p$ = 114 meV in zero magnetic field.

Now let us consider the ac electrical transport. Fig. 8 shows the temperature dependence of the ac resistance ($R$) and reactance ($X$) of $Pr_{0.6}Sr_{0.4}MnO_3$ for three frequencies, $f$ = 1, 5 and 10 MHz measured in zero and low dc magnetic fields ($H$ = 300, 500, 700, and 1 kOe). When $f$ = 1, 5, a small step like increase in $R$ occurs around $T_C$ = 304 K. The feature is more pronounced for $f$ = 10 MHz. We can also see a step-like decrease at $T_S$ = 86 K, which closely correlates with the magnetization data. The applied magnetic field decreases the magnitude of anomaly at $T_C$ and at $T_S$. Under a magnetic field of $H$ = 1 kOe, the step-like increase at $T_C$ is completely suppressed leading to ac magnetoresistance of -4%. This value is an order of magnitude smaller than the ac magnetoresistance near $T_C$ found in $La_{0.7}Sr_{0.3}MnO_3$.[32] In contrast to the ac resistance, $X(T)$ in zero field shows clear anomalies both $T_C$ and $T_S$ even for $f$ = 1 MHz. The anomalies are clearly suppressed with increasing dc magnetic field. There is a qualitative change in the behavior of $X(T)$ at $f$ = 10 MHz. Below 200 K, $X(T)$ raises with lowering temperature unlike for $f$ = 1 and 5 MHz. In the absence of theoretical model of ac electrical transport, the features observed in the ac resistance and the reactance can be only qualitatively explained as follows. The flow of ac current through the sample creates an oscillating circular magnetic field transverse to the direction of the current. The ac magnetic field produced is roughly 20 mOe. At low frequency, the current flows through the bulk of the sample but as the frequency increases current flow tends to migrate towards surface. High frequency current flows in a shell of skin depth thickness $\delta = (1/\pi f \mu_0 \mu_t)^{1/2}$ from the top and bottom surfaces of the sample.

The ac impedance of the sample is $Z(f, T, H) = R(f, T, H) + iX(f, T, H)$ where $R$ is the ac resistance and $X$ is the reactance. The reactance $X = 2\pi f L$ is due to the self inductance ($L$) of the sample which is related to the ac transverse permeability of the sample ($L=G\mu_t$ where G is the geometrical factor and $\mu_t$ is the transverse permeability). When the skin depth is larger than thickness of the sample and current flow is uniform in the sample. The circular ac magnetic field created by the ac current interacts with the magnetization of the sample. At the onset of ferromagnetic transition, the $\mu_t$ in zero external magnetic field increases rapidly which causes the $L$, hence the $X(T)$, to increase abruptly at $T_C$. The decrease of $X(T)$ around $T_S$ = 86 K indicates the decrease of $\mu_t$ due to the structural phase transition. In the presence of an axially applied dc magnetic field, $\mu_t$ is expected to decrease since the small ac magnetic field becomes inefficient to rotate the magnetization away from the direction of the dc magnetic field. Hence, the maximum decrease of $\mu_t$ occurs at temperatures just below the $T_C$. Hence, the $X(T)$ just below $T_C$ decreases with increasing $H$. As the frequency increases, the skin depth also decreases which causes the ac resistance ($R$) to increase since the available cross sectional area for the current flow decreases. Since $\delta \propto 1/\sqrt{(f\mu)}$, $R$ increases abruptly at $T = T_C$ and also shows a feature at $T= T_S$ when $f$ = 10 MHz The decrease in skin depth also affects the behavior of the reactance. The transverse permeability decreases and hence $\delta$ increases with increasing strength of the dc magnetic field, which results in suppression of the ac resistance near $T_C$. The magnitude of ac magnetoresistance depends on the resistivity and transverse permeability of the sample. Due to high value of transverse permeability ($\mu \approx 10^4$) and low resistivity ($\rho$ = 100- 400 $\mu\Omega$ cm), amorphous and nanocrystalline alloys show much larger ac magnetoresistance (= 70-90 % for H = 10-100 Oe) than manganites but it saturates in kOe field.[33]

(D). Thermoelectric power in zero and non zero magnetic field

Fig.9 shows the temperature dependence of thermoelectric power ($Q$) measured under H = 0 and 3 T. The $Q$ is negative ($\approx$ -20 $\mu$V/K) at 360 K and it smoothly decreases ($Q$ becomes less negative) as $T_C$ is approached from the high temperature side. A rapid change in $Q$ occurs around $T_C$, and then $Q$ decreases smoothly towards zero in the ferromagnetic state as $T$ approaches 90 K. Around $T_S$ = 86 K, $Q$ shows a dip and then it increases slowly. It is to be noted that the rapid change in $Q$ occurs at $T_C$ rather than at the temperature where the dc resistivity shows a maximum ($T_p$) which clearly indicates that thermopower probes the intrinsic property of the grain and insensitive to grain boundaries. In electrical transport measurement, voltage drop across the high resistive grain boundary is also sampled along with the voltage drop in the low resistance ferromagnetic grains. Because no current flows, thermopower of individuals grains is additive, independent of the resistivity of intergranular connections and hence $Q$ probes the intrinsic phase transition within grains. An applied external magnetic field of $H = 3$ T eliminates the anomaly around the $T_C$ but hardly affects $Q$ much below 250 K, unlike the influence of $H$ on $\rho(T)$.

When the electrical transport in the paramagnetic state ($T > T_C$) is dominated by thermally activated hopping of polarons, $Q$ is expected to obey the relation

$$Q = \frac{k_B}{e}\left(\frac{E_s}{k_B T} + \alpha'\right) \quad \ldots\ldots\ldots\ldots (4)$$

where $k_B$ is Boltzmann constant, $e$ is electronic charge, and $\alpha'$ is a sample dependent constant related to the kinetic energy of polarons.[34] If $\alpha' < 1$, the transport follows the small polaron hopping model and for $\alpha' > 2$ it follows the large polaron hopping.[35] We show $Q$ versus $1/T$ and the fit based on the above relation in the inset (b) of fig.7. The calculated values of $\alpha'$ is 0.515 for 0 T is less than 1. The calculated activation energy for $Q$ is $E_s = 26$ meV which is much lower than the activation energy ($E_\rho = 117$ meV) obtained from the electrical transport. Such a large difference between activation energies for electrical resistivity and thermopower is considered to be a hallmark of polaronic transport above $T_C$. The $Q$ is negative in the temperature range from 30 to 360 K, which suggests that the charge carries are predominantly electron-like at the Fermi level. Negative value of $Q$ over a wide temperature was also found for x = 0.48-0.55 in $Pr_{1-x}Sr_xMnO_3$ series.[15] The composition with x < 0.5 are supposed to be hole doped, however it is not uncommon to find negative $Q$ for x < 0.5. For example, $Q$ in $La_{1-x}Ca_xMnO_3$ series shows negative value at room temperature even for x = 0.3 or Q shows a change of sign as a function of temperature.[36,37,38] The sign of $Q$ in manganites is affected by contribution from spin disorder term ($Q_s$ = -20 µV/K) in the paramagnetic state and carrier entropy due to presence of correlated polarons or charge ordered nanoclusters and these make the analysis difficult unless the measurement is extended to very high temperature.[29,30,39] We are more interested in the change of $Q$ with the magnetic field rather than the value and sign of $S$ in zero field alone. The magnetothermopower defined as MTEP = $[Q(H)-Q(0)]/Q(0)$ is shown on the right scale. The MTEP is zero at 360 K and it increases in magnitude with lowering temperature, goes through a peak around $T_C$ and then decreases to zero above 250 K. The value of MTEP reaches 25% at the $T_C$ f or $\Delta H$ = 3 T. The observed value of MTEP is comparable to the reported value 38% for $\Delta H$ = 5 T at $T_C$ = 225 K in $La_{0.67}Ca_{0.33}MnO_3$ film by Boxing Chen et al.[40] Available literature on MTEP in manganites are very few compared to magnetoresistance.[41,42,43,44]

We have also measured the field dependences of $Q$ and $\rho$ at selected temperatures ($T$ = 290-320 K with $\Delta T$ = 5 K). We plot the MTEP and MR as a function of $H$ in Fig. 10 (a) and (b), respectively. The isotherms at $T$= 305 and 310 K showed a large change in $Q$ and $\rho$ as the field is swept from $H$ = 0 to 3 T. No hysteresis was observed while reducing the field to zero. In order to seek a correlation between the MR and the MTEP, we plot [-$\Delta Q/Q(0)$] against [-$\Delta \rho/\rho(0)$] for the above temperatures in Fig. 10(c). The curves for different temperatures for $T \geq T_C$ falls on each other suggesting that the MR and MTEP could share a common mechanism. The MTEP and the MR are linear at low fields above and below $T_C$. Both MTEP and MR reach a maximum at $T = T_C$. It is interesting to note that the maximum $\Delta Q/Q(0)$ at $T_C$ reaches 25% which is larger than the dc magnetoresistance ($\Delta \rho/\rho$ = 14%) for $\Delta H$ = 3 T. A close connection between MTEP and MR is expected in Mott's expression for $Q$. According to the Mott's model, diffusive component of $Q$ is dependent on the energy dependent of the resistivity at the Fermi level.[45]

$$Q = \frac{\pi^2}{3} \frac{k_B T}{e} \frac{\rho'(E_F)}{\rho(E_F)}$$

where $e$ is the elementary charge of the carrier, and $\rho'(E_F) = \partial \rho(\varepsilon)/\partial \varepsilon |_{E=E_F}$. If we assume $\rho'(E_F)$ = constant, then $\Delta Q/Q \propto \Delta \rho/\rho$. Magnetoresistance around $T_C$ in manganites originates not only from the suppression of spin fluctuations but also from the magnetic field dependence of spin-phonon coupling as suggested by magnetothermal conductivity study.[46] Spin-phonon coupling originates from presence of lattice polarons that form when $e_g$ electron in the paramagnetic state localizes on a $Mn^{3+}$ site inducing a local Jahn-Teller distortion of the $MnO_6$ octahedra. The lattice polarons in manganites also have magnetic cloud around them (known as magnetic polaron) and they form at temperature $T \approx 1.8\ T_C$. They grow in density and size as the $T_C$ is approached as suggested by combined studies of thermal expansion and small

angle neutron scattering under a magnetic field in Y-doped $La_{0.67}Ca_{0.33}MnO_3$.[47] The application of magnetic field near the $T_C$ causes the magnetic polarons to coalesce and in that process $e_g$ electron becomes delocalized. This leads to a large negative MR and it may also contribute to the observed MTEP apart from field-induced suppression of spin fluctuations. A close correlation between MTEP and MR was also seen in granular giant magnetoresistive granular alloys [48] and more recently in magnetic tunnel junction, where MTEP is connected with the asymmetry in the density of states for spin up and down bands.[49]

In high thermoelectric oxides such as layered $Na_xCoO_2$[50] and misfit cobalt oxide $[Bi_{1.7}Co_{0.3}Ca_2O_4]^{RS}_{0.6} CoO_2$,[51] the origin of MTEP was suggested to suppression of the spin entropy of $Co^{3.5+}$ ions under a magnetic field, and a scaling relation between $\Delta Q(H)/Q$ and $\mu_0 H/T$ was demonstrated. We have also attempted such a scaling in our compound. In Fig. 11(a) we plot $\Delta Q(H)/Q$ versus $H/T$. However, we fail to observe a scaling behavior in our sample which suggest that there are additional contributions other than spin entropy alone to the observed MTEP. In a simplistic view, thermopower is measure of entropy per carrier ($Q = -\frac{\sigma}{Ne}$ where $\sigma$ is the total entropy and $N$ is the carrier concentration) Hence, we look for possible correlation between changes in the magnetic entropy ($-\Delta S_m = [S_m(H)-S_m(0)]$) and thermopower [$\Delta Q = Q(H)-Q(0)$]. In Fig. 11(b) we plot, $\Delta Q$ versus $-\Delta S_m$. It is found that the curves at different temperatures are nearly linear though they don't coalesce. It should be noted that the estimated entropy change from applying the Maxwell relation to the magnetization isotherms is not just the spin entropy ($S_m$) alone but it is sum of contributions from lattice ($S_{lattice}$), charge carrier entropy ($S_e$), spin entropy ($S_m$). Generally, $S_{lattice}$ is taken to be independent of magnetic field in a second order PM-FM transition. Since this compounds shows appreciable magnetoresistance, the charge carrier entropy is also affected with the magnetic field besides the spin entropy. Since we have seen a close correlation between the magnetothermopower and magnetoresistance in Fig. 10 (c), the magnetoresistance might have a connection with the magnetic entropy change as well. In Fig. 11(c), we plot the resistivity change in different magnetic fields, i.e, $-\Delta\rho = [\rho(H)-\rho(0)]$ versus the magnetic entropy change ($-\Delta S_m$) at different temperatures. From the figure, it is clear that $-\Delta\rho$ changes nearly linearly with $-\Delta S_m$ at all the measured temperatures. Hence, all these three quantities, $\Delta Q$, $\Delta S_m$ and $\Delta\rho$ appear to be interrelated most likely through their dependences on the intrinsic magnetization. Further research on other manganites and other compounds is highly desirable for comparison as well as to understand a common mechanism relating magnetothermopower, magnetoresistance and magnetic entropy change.

## IV. Summary

In summary, we have investigated magnetization, direct and alternating current magnetoresistance, magnetocaloric effect and magnetothermopower in $Pr_{0.6}Sr_{0.4}MnO_3$ sample having ferromagnetic transition just above room temperature ($T_C$ = 305 K). Magnetization upon cooling showed a sharp decreases at $T_S$ = 86 K within the long range spin ordered state and this feature was accompanied by and exothermic peak in differential thermal analysis. This low temperature anomaly in magnetization was attributed to orthorhombic to monoclinic structural transition while cooling. The structural transition leads to inverse magnetocaloric effect at $T_S$ whereas normal magnetocaloric effect peaks around $T_C$. Coexistence of a large $\Delta S_m$ value of 3.416 J/Kg K and refrigeration capacity of 103.34 J/Kg for $\Delta H$ = 3 T and magnetothermopower (=25 % for $\Delta H$ = 3 T) make this compound interesting for applications for room temperature magnetic refrigeration, magnetically tunable thermoelectric power generators and heat pumps. In addition, we have found a close correlation between magnetoresistance, magnetothermopower and magnetic entropy change in the same compound. We have also shown that ac electrical transport enacts the behavior of the ac susceptibility and it provides a simple means of investigating interplay between charge transport and magnetism simultaneously.

**Acknowledgements:** R. M. Acknowledges the National Research Foundation, Singapore for supporting this work (No. NRF-CRP-G-2007-12)

**Figure Captions:**

**Fig.1.** Temperature dependence of the dc magnetization($M$) for $Pr_{0.6}Sr_{0.4}MnO_3$ under different magnetic fields. The inset **(a)** shows exothermic and endothermic peaks in a temperature cycle. $T_S$ is the structural transition temperature.

**Fig.2.** Isothermal magnetization $M(T,H)$ at 10 and 120 K for $Pr_{0.6}Sr_{0.4}MnO_3$. The inset **(a)** shows isothermal $M$-$H$ curves from 10-360 K and **(b)** $M$-$H$ curves for 10-180 K. A crossover behavior in $M(H)$ occurs below $T = T_S$.

**Fig.3.** **(a)** $M(H)$ isotherms around $T_C$ (T= 294-318 K with $\Delta T = 2$ K); **(b)** Arrott-plots ($M^2$ versus $H$) following mean field theory ($\beta = 0.5$ and $\gamma = 1$); **(c)** Modified Arrott plots ($\beta = 0.365$ and $\gamma = 1.336$); **(d)** Spontaneous magnetization $M_S(T)$ (left) and inverse initial susceptibility $\chi_0^{-1}$(right) were fitted by equations (1) and (2); **(e).** Isothermal $M(H)$ near $T = T_C$ in log-log scale and solid lines are the linear fitting by Eq. (3)

**Fig.4.** Magnetic entropy change ($\Delta S_m$) as a function of temperature for different $H$. The inset **(a)** shows ($\Delta S_m$) (left scale) and refrigeration capacity (RC) as a function of magnetic field (right scale).

**Fig.5.** (Main panel) Temperature dependence of the exponent ($n$) in the power law dependence $\Delta S_m(H) = H^n$. The insets show the field dependences of $S_m$ at $T > T_C$, at $T = T_C$ and $T < T_C$.

**Fig.6.** Scaling of normalized $\Delta S_m /\Delta S_{Max}$ versus $\Theta = (T-T_p)/(T_r-T_p)$ curves for different magnetic fields.

**Fig.7.** Temperature dependence of resistivity($\rho$) under H= 0, 3T (left scale) and percentage of negative magneto resistance (-%MR) (right scale).

**Fig.8.** Temperature dependence of the ac resistance ($R$) and reactance ($X$) for selected frequencies (f = 1, 5 and 10 MHz) under different dc magnetic fields ($H$ = 0, 300, 500, 700 Oe and 1 kOe).

**Fig.9.** Temperature dependence of thermoelectric power ($Q$) (left scale). The right scale shows magnetothermopower(MTEP). The inset shows the data and the polaronic fit for $Q$ vs $1/T$ at $T>T_C$

**Fig.10.** (a) Field dependence of the magnetothermopower ($\Delta Q/Q$) and (b) magnetoresistance ($\Delta\rho/\rho$) at selected temperatures around $T_C$. (c) Correlation between magnetothermpower and magnetoresistance at $T = 320, 315, 310$ K $> T_C$ and $T = 310, 305, 300, 295$ and $290$ K $< T_C$.

**Fig.11.** (a) $\Delta Q/Q$ versus $H/T$, (b) $\Delta Q$ versus $-\Delta S_m$ and (c) $-\Delta\rho$ versus $-\Delta S_m$ at selected temperatures around $T_C$.

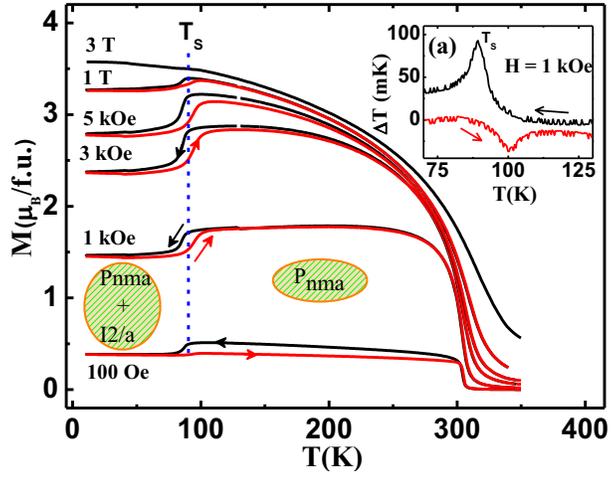

**Figure 1**

D.V. Maheswar Repaka *et.al.*

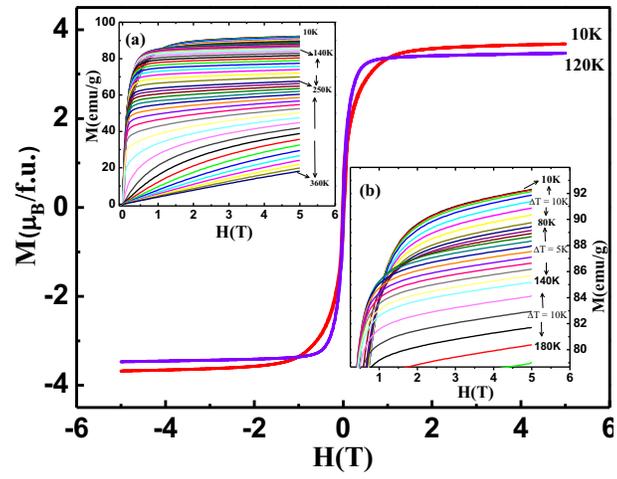

**Figure 2**

D.V. Maheswar Repaka *et.al.*

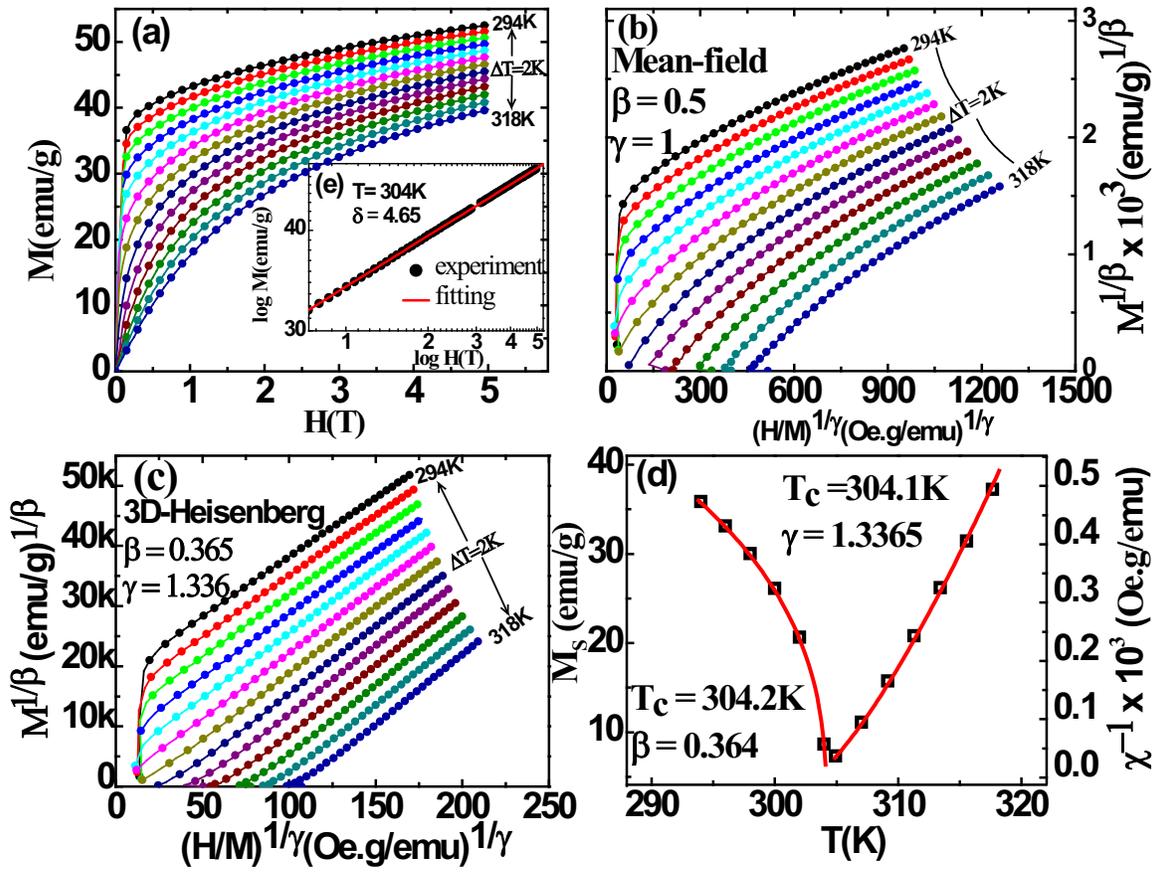

Figure 3

D.V. Maheswar Repaka *et.al*.

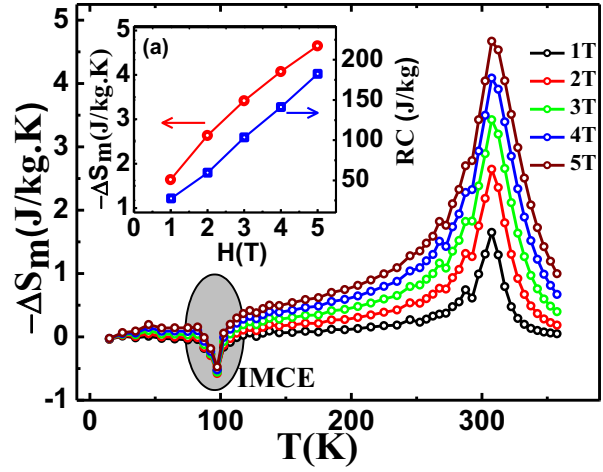

**Figure 4**

D.V. Maheswar Repaka *et.al.*

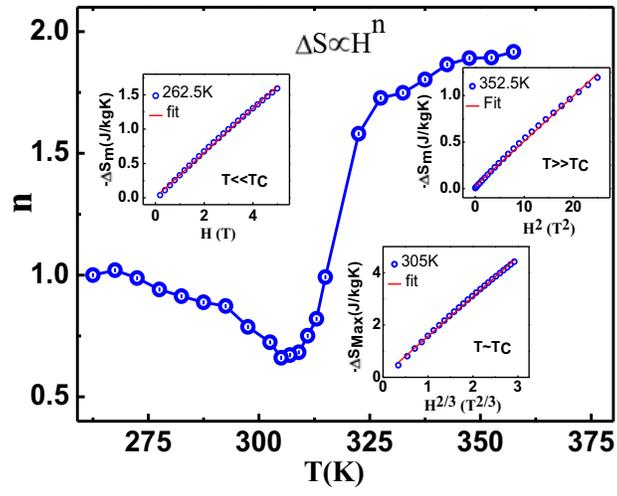

**Figure 5**

D.V. Maheswar Repaka *et.al.*

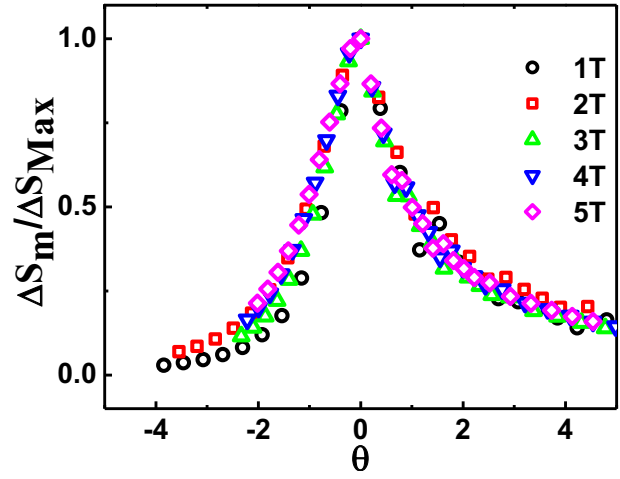

**Figure 6**

D.V. Maheswar Repaka *et.al.*

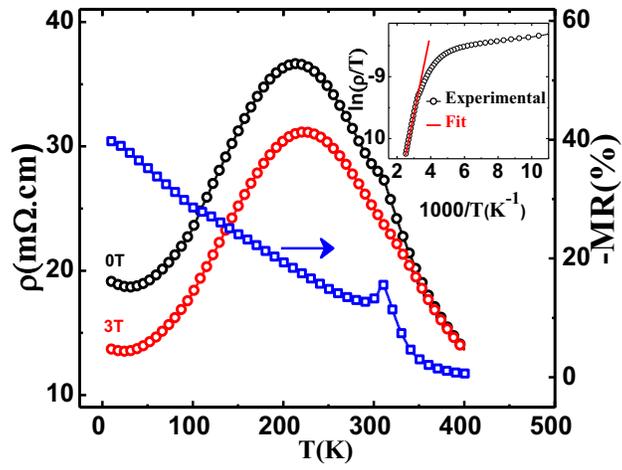

**Figure 7**

D.V. Maheswar Repaka *et.al.*

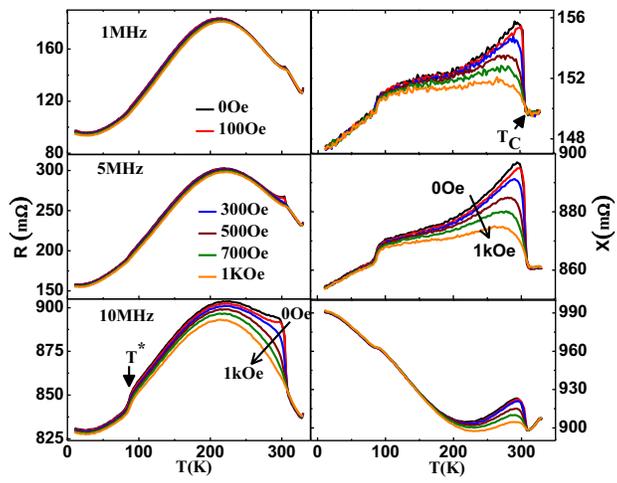

**Figure 8**

D.V. Maheswar Repaka *et.al.*

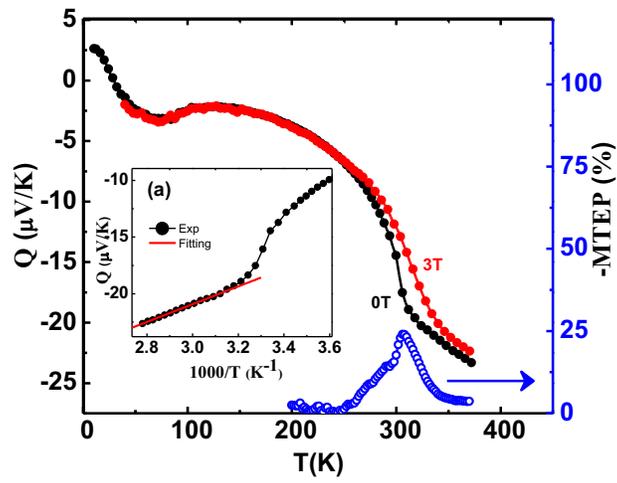

**Figure 9**

D.V. Maheswar Repaka *et.al.*

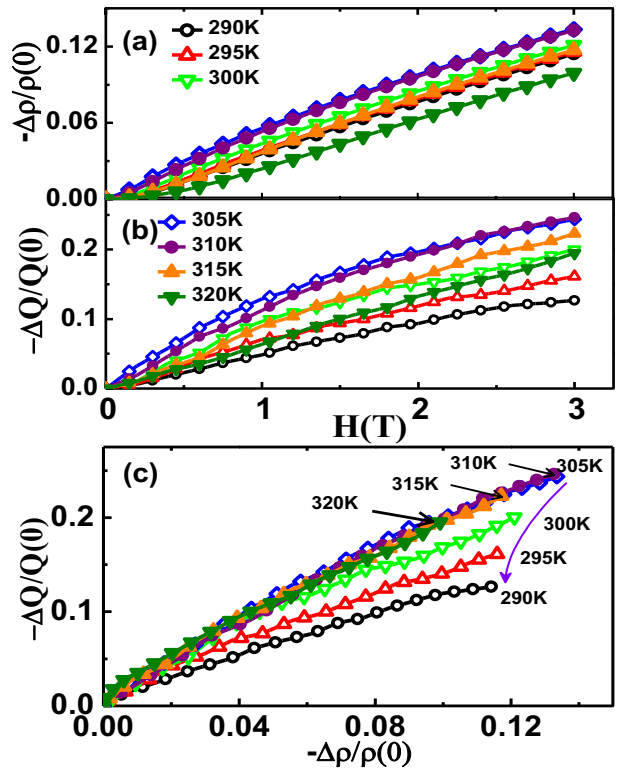

**Figure 10**

D.V. Maheswar Repaka *et.al.*

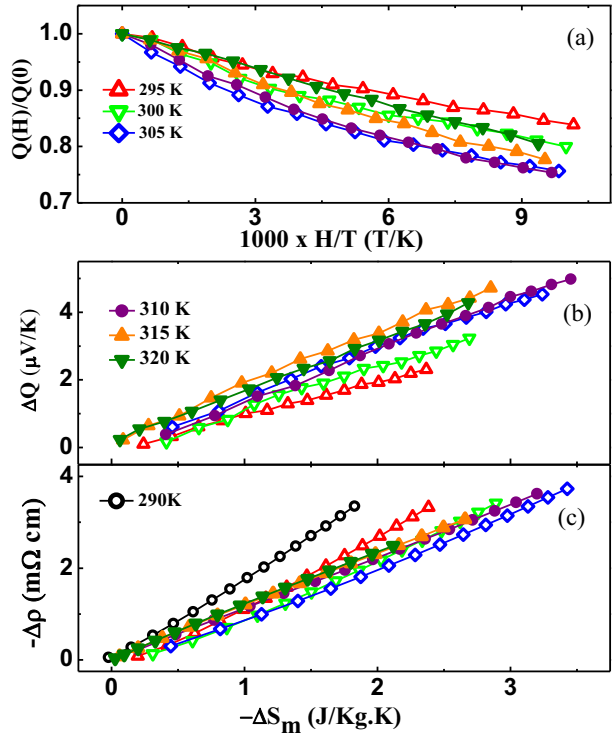

**Figure 11**

D.V. Maheswar Repaka *et.al.*